\documentclass[aps,pra,twocolumn,superscriptaddress,numerical]{revtex4-1}
\usepackage{bm}
\usepackage{hyperref}
\usepackage[english]{babel}
\usepackage[T1]{fontenc}
\usepackage{graphics}
\usepackage{bbm}
\usepackage{amsthm}
\usepackage{amsmath}
\usepackage[amssymb,Gray]{SIunits}
\usepackage{xcolor}
\usepackage{mathtools}
\usepackage{amsfonts}
\usepackage{mathpazo}
\usepackage{braket}
\usepackage[parfill]{parskip}

\newcommand{\abs}[1]{\ensuremath{\left\vert #1 \right\vert}}

\newcommand{\ve}{\mathbf}
\DeclarePairedDelimiter{\norma}{\lVert}{\rVert}

\begin{document}


\title{Collapse in ultracold Bose Josephson junctions}

\author{M. Bilardello}
\email{marcomaria.bilardello@phd.units.it}
\affiliation{Department of Physics, University of Trieste, Strada Costiera 11, 34014 Trieste, Italy}
\affiliation{Istituto
Nazionale di Fisica Nucleare, Trieste Section, Via Valerio 2, 34127 Trieste,
Italy}

\author{A. Trombettoni}
\email{andreatr@sissa.it}
\affiliation{ CNR-IOM DEMOCRITOS Simulation Center, Via Bonomea 265, 
I-34136 Trieste, Italy}
\affiliation{SISSA, Via Bonomea 265, I-34136 Trieste, Italy}
\affiliation{Istituto
Nazionale di Fisica Nucleare, Trieste Section, Via Valerio 2, 34127 Trieste,
Italy}

\author{A. Bassi}
\email{bassi@ts.infn.it}
\affiliation{Department of Physics, University of Trieste, Strada Costiera 11, 34014 Trieste, Italy}
\affiliation{Istituto
Nazionale di Fisica Nucleare, Trieste Section, Via Valerio 2, 34127 Trieste,
Italy}

\date{\today}

\begin{abstract}
We investigate how ultracold atoms in double well potentials 
can be used to study and put bounds on models describing wave function 
collapse. We refer in particular to the continuous spontaneous localization 
(CSL) model, which is the most 
well studied among dynamical reduction models. It modifies 
the Schr\" odinger equation in order to include the collapse 
of the wave function in its dynamics. 
We consider Bose Josephson junctions, where ultracold bosons 
are trapped in a double well potential, since 
they can be experimentally controlled with high accuracy and are 
suited and used to study 
macroscopic quantum phenomena on scale of microns with a number of particles 
typically ranging from $\sim 10^2-10^3$ to $\sim 10^5-10^6$. 
We study the CSL dynamics of three atomic states showing macroscopic 
quantum coherence: the atomic coherent state, the superposition of 
two atomic coherent states, and the NOON state. We show that 
for the last two states 
the suppression of quantum coherence induced by CSL model increases 
exponentially with the number of atoms. We observe that, 
in the case of optically trapped atoms, the spontaneous photon emission 
of the atoms induce a dynamics similar to the CSL one and we conclude 
that magnetically trapped atoms may be more convenient 
to experimentally test the CSL model. We finally discuss 
decoherence effects in order to provide reasonable 
estimates on the bounds that it is (or it will) possible to 
obtain for the parameters of the CSL model in such class of experiments:  
as an example, we show that a NOON state with $N \sim 10^3$ with a coherence 
time of $\sim 1$ s can constrain the CSL parameters in a region where 
the other systems presently cannot.
\end{abstract}


\maketitle


\section{Introduction}\label{sec:intro}

Matter-wave optics and atom interferometry are nowadays routinely 
used for a variety of precision measurements~\cite{rmp_atom_interferometry}, 
paving the way to the development of the tools needed  
to perfom a class of experiments aimed at testing quantum mechanics and 
general relativity, with applications ranging from metrology to 
atomtronics~\cite{focus_1,focus_2}. In this paper we focus on the use of ultracold atoms to study wave function collapse and we consider atomic gases weakly coupled via a double well potential. Such a setup provides an atomic counterpart of Josephson devices~\cite{barone,cataliotti} and it is has been experimentally realized and studied for both ultracold bosons~\cite{Alb05,Lev07,LeB11,trenkwalder2016quantum} and  fermions~\cite{Val15}. A Josephson physics and coherent tunneling 
have been also studied not only in space (as in double well potential) but also between internal levels~\cite{Sme03,Zib10,Gro10,Rie10}. Due to the high tunability  of experimental parameters, Bose Josephson junction is one of the paradigmatic setups in which to probe and study quantum coherence on mesoscopic/macroscopic scale. For this reason, we use it to study the bounds that can be put on models for the wave function collapse. 

The Continuous Spontaneous Localization (CSL) model~\cite{ghirardi1990markov} is the most well studied model among dynamical reduction models~\cite{bassi2003dynamical,bassi2013models}. It adds extra terms to the Schr\" odinger equation in order to describe the collapse of the wave function. The new terms are stochastic and non-linear to mimic the 
collapse while avoiding faster-than-light signaling~\cite{gisin1989stochastic,gisin1990weinberg,gisin1995relevant}. 
The resulting modified Schr\" odinger equation preserves the quantum mechanical predictions for microscopic systems, 
while inducing macroscopic system to be localized in space. 

The CSL model contains two new phenomenological parameters: 
a collapse rate $\lambda$ and a resolution length $r_C$. 
The differences between standard quantum mechanics and the 
CSL model depend on the values of these two new parameters. 
With too small values, CSL predictions are practically indistinguishable from the quantum mechanical ones 
(the model then looses its effectiveness); on the other extreme, too large values are excluded since then CSL would contradict known experimental facts. Ghirardi, Rimini and Weber proposed a collapse rate 
$\lambda= 10^{-16}$ s$^{-1}$ and a resolution length $r_C = 10^{-7}$ m 
in their first dynamical reduction model~\cite{ghirardi1986unified}, 
slightly modified later with $\lambda= 10^{-17}$ s$^{-1}$ 
(and same resolution length) for the CSL model~\cite{ghirardi1990markov}. 
Adler proposed stronger values, $r_C = 10^{-7}$ m and 
$\lambda = 10^{-8 \pm 2}$ s$^{-1}$, and $r_C = 10^{-6}$ m and 
$\lambda = 10^{-6 \pm 2}$ s$^{-1}$, based on the analysis of the process 
of latent image formation in photography~\cite{adler2007lower}.

Bounds on the parameters are of course ultimately set by experiments. 
In the last years, a great effort in comparing collapse models 
with experiments has been made, through a large variety of experimental 
setups: from matter-wave interferometry~\cite{torovs2016bounds}, 
to heating effects in cantilevers~\cite{PhysRevLett.116.090402,vinante2016improved}, 
to spontaneous X-rays emission from matter~\cite{curceanu2015spontaneously}, 
which currently sets the strongest upper bound 
$\lambda \leq 10^{-11}$ s$^{-1}$ (for $r_C = 10^{-7}$ m). The resulting exclusion plot in the $\lambda - r_C$ parameters space is shown in fig.~\ref{excplot}, where the white region must still be experimentally probed, while the non-white regions have been excluded by experiments listed before. Taking into account the challenge to entirely probe this white region by currently used experimental setups, and considering that on-going experiments with atomic gases in double well potentials are able to create and detect strongly correlated entangled states~\cite{Sch16}, we think is of great interest to study the possibility to perform experimental tests of quantum mechanics using ultracold bosons in Bose Josephson junctions and to determine what requests the experimental values have to satisfy in order to put new bounds on the parameters of the CSL model.

Although the spontaneous localization mechanism of collapse models 
is very weak, the progresses in the last two decades in the control and 
manipulation of ultracold gases make such system a promising setup 
to test the CSL model. 
Together with their very low temperature ($T \sim 10-100$ nK), ultracold 
atomic systems are characterized by quantum phenomena 
involving a mesoscopic/large number of atoms ($N$ ranging from 
$10^{2}-10^3$ to $10^{5}-10^6$)~\cite{bloch2008many,pitaevskii2016bose}. 
A first work that studied the CSL effects in a cold-atomic system ~\cite{laloe2014heating} set the upper bound of $\lambda \leq 10^{-7}$ s$^{-1}$ (for $r_C = 10^{-7}$ m) from the analysis of the heating effects on a Bose Einstein condensate. 
In~\cite{Bil16} a theoretical analysis of a recent experiment~\cite{Kov15} was performed, where an out-of-equilibrium 
gas of $^{87}Rb$ atoms was cooled to a temperature $T \sim 50$ pK. The dynamical equations 
for the CSL model were studied, concluding that the resulting bounds are beaten only by measurements of the spontaneous X-ray emission and by experiments with cantilevers. It was as well shown that the bounds are not changed by non-Markovian extensions of the CSL model.

\begin{figure}[t]
\label{double}
\includegraphics[scale=0.53]{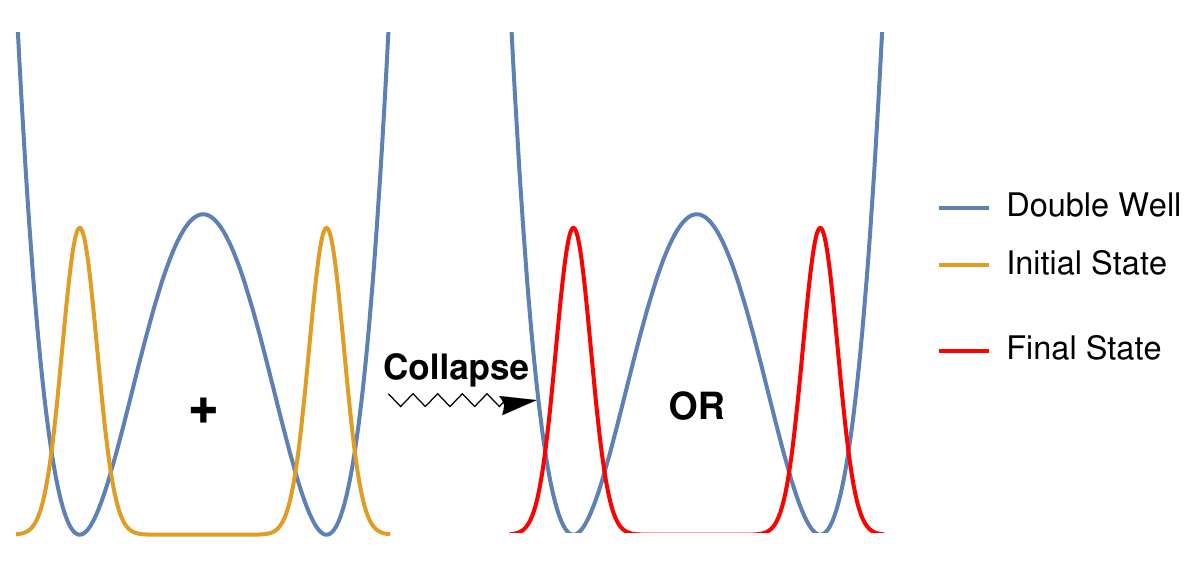}
\caption{Action of the collapse noise on an atom in a double well
potential (blue curve). The initial state (yellow curve) is
delocalized over the two wells. The interaction of the atom with the
collapse noise localizes the atomic state in one of the two wells (red
curve).}
\end{figure}

In this paper we study how CSL affects cold atoms in a double well potential, i.e., a Bose Josephson junction, as qualitatively depicted in fig.~\ref{double}, where we also schematically describe the effect of the collapse of the wave function. 
A Bose Josephson junction can be experimentally implemented \textit{e.g.} by superimposing 
an optical lattice to a parabolic potential~\cite{Alb05,Mor06} or by the use of a laser beam to create the barrier~\cite{Val15}. 
In a Bose Josephson junction, when the barrier is high enough (much larger 
than the chemical potential), the atoms can be only in two states. 
Each wave function is spatially localized in one of the two wells: 
the higher the barrier separating the two wells, the 
lower the overlap between the two wave functions. 
Experimental quantities such as the height of the barrier and the geometry of each well~\cite{pitaevskii2016bose}, can be tuned with a high degree of control. 

A phase difference between the two wells emerges as a consequence of the weak 
link also for a small number of atoms~\cite{Buc13} and it can be fixed (giving rise to the so-called phase state). 
This possibility for example leads to use Bose Josephson junctions as an atom 
interferometer~\cite{shin2004atom,schumm2005matter,rmp_atom_interferometry}. 
The quantum mechanical evolution of the phase state is characterized 
by the typical collapse and revival of the interference fringes, 
while the CSL dynamics decreases the interference until 
the fringes disappear and each atom is localized in one of the two wells. In this paper we will determine how fast the CSL dynamics collapses phase states, compared to the typical longest coherence time experimentally detected (\textit{e.g.} $\approx 200$ ms in~\cite{jo2007long}).

Besides the phase states, Bose Josephson junctions provide as well 
a promising possibility for creating macroscopic entangled states, 
including Schr\" odinger's cat 
states~\cite{gordon1999creating,huang2006creation}. The use of squeezed states 
in atomic interferometers and clocks leading to sub-shot-noise 
performance has been 
exploited in experiments~\cite{Gro10,Rie10,Luc11,Kru16}, while the 
detection of Bell correlations between the spins of $\sim 500$ atoms 
in a Bose-Einstein condensate was recently reported in~\cite{Sch16}. 
In this paper, we focus on two types of macroscopic entangled states: 
the superposition of two phase states and the so-called 
NOON state. Even if a Schr\" odinger's cat state has 
not been experimentally detected yet in a double well potential, 
for these two macroscopic entangled states several  
preparation techniques have been proposed and 
analyzed \cite{Cir98,Dal00,Dun01,louis2001macroscopic,Maz07} 
(see a review in \cite{ferrini2012macroscopic}). 

Macroscopic entangled states are of notoriously extremely sensitive to external noise sources. Therefore in this paper 
we will study the collapse induced by CSL dynamics to the macroscopic 
entangled states considered here, comparing the result with two typical decoherence sources in optically trapped systems: 
phase noise~\cite{PhysRevA.84.043628} and spontaneous photon emission process~\cite{cohen1977frontiers,cohen1990atomic,pichler2010nonequilibrium}. 
The effects of thermal fluctuations and three body recombination terms will be also addressed. 

The paper is organized as follows: we start by briefly describing the Bose Josephson junction system 
within the two mode approximation and introducing the coherent atomic state (or phase state) 
and the two macroscopically entangled states we consider
in our investigation. We then introduce the CSL model, and we find the 
density matrix evolution for a gas of bosonic atoms in a Bose Josephson 
junction with CSL dynamics. Next, we find the correlation evolution 
for the phase state, for the superposition of phase states and for the NOON 
state. Finally, we compare the CSL dynamics with several 
typical decoherence sources: phase noise, 
spontaneous photon emission process, thermal effects and three body 
recombination terms. 
A discussion of the results and of perspectives is presented in the 
conclusions.
%

\section{Two-mode model for the Bose Josephson junction}\label{sec:bjj}

Within the usual two mode approximation~\cite{sme97}, valid for large energy barriers 
between the two wells (and much larger than the chemical potential), 
one assumes that the atoms can be either in the state 
$\ket{\psi}_L$ of the left well, or in the state $\ket{\psi}_R$ 
of the right well. The left and the right states 
are taken orthogonal $\braket{\psi_R|\psi_L} = 0$ (see~\cite{sme97}). 
In the second quantization formalism, 
the Hamiltonian for the interacting gas is
\begin{equation}
\label{hamiltonian}
\begin{split}
\hat{H} = &\int d\ve{x} \, \hat{a}^{\dagger} (\ve{x}) \left (-\frac{\hbar^2 \nabla^2}{2m} +V_{\text{DW}}(\ve{x}) \right) \hat{a}(\ve{x}) + \\
&\frac{g}{2} \int d\ve{x} \,  \hat{a}^{\dagger} (\ve{x}) \hat{a}^{\dagger} (\ve{x}) \hat{a}(\ve{x}) \hat{a}(\ve{x}),
\end{split}
\end{equation}
where $g = 4 \pi \hbar^2 a_S /m$ is the coupling constant of the atom-atom interaction, with $a_S$  the scattering length of the atoms, and we introduced the external double well potential $V_{\text{DW}}$ whose specific form 
determines the left and right Wannier wave functions $\psi_L (\ve{x})$, $\psi_R (\ve{x})$ (and therefore the 
coefficients of the two-mode model). The total number of particles is fixed to $N$. 
%
Usually, the wave functions $\psi_L (\ve{x})$, $\psi_R (\ve{x})$ are well approximated from the solutions of the time-independent 
Gross-Pitaevskii equation:
\begin{equation}
\label{gpe}
-\frac{\hbar^2 \nabla^2}{2m}\psi \left( \ve{r}\right) + V_{\text{DW}}\left( \ve{r}\right)\psi \left( \ve{r}\right) + g\left \lvert\psi \left( \ve{r}\right)\right \rvert^2 \psi \left( \ve{r}\right)= \mu \psi \left( \ve{r}\right).
\end{equation}
Denoting by $\psi_{\text{G}}$ and $\psi_{\text{E}}$ the ground and the first-excited 
states of~\eqref{gpe}, then one has $\psi_L (\ve{x})=\left( \psi_{\text{G}} (\ve{x}) 
+ \psi_{\text{E}} (\ve{x}) \right)/\sqrt{2}$ 
and $\psi_R (\ve{x})=\left( \psi_{\text{G}} (\ve{x}) - \psi_{\text{E}} (\ve{x}) \right)/\sqrt{2}$. This way of constructing 
the wave functions $\psi_L (\ve{x})$, $\psi_R (\ve{x})$ is typically good if the number of particles per well is not 
too large and they do not depend on the interactions via the atom numbers $N_L$ and $N_R$ (in that case 
one has to resort to a nonlinear tight-binding ansatz~\cite{Sme03_bis,Jez13}).

Let us introduce $\hat{a}^{\dagger}_{L} \left(\hat{a}_{L}\right)$ and $\hat{a}^{\dagger}_{R} \left(\hat{a}_{R}\right)$ 
the creation (annihilation) operators for, respectively, the left and right states. Rewriting the 
operators $\hat{a} (\ve{x})$ in terms of $\hat{a}_{L}$ and $\hat{a}_{R}$ as
\begin{equation}
\label{creationtwomode}
\hat{a} (\ve{x}) = \psi_L (\ve{x}) \hat{a}_L +\psi_R (\ve{x}) \hat{a}_R
\end{equation}
(with the wave functions $\psi_L (\ve{x})$, $\psi_R (\ve{x})$ appropriately normalized to $1$), 
the Hamiltonian operator in eq.~\eqref{hamiltonian} becomes~\cite{sme97,javanainen1999splitting}
\begin{equation}
\label{hamiltoniantwomode}
\hat{H} = -J \left ( \hat{a}_{L}^{\dagger} \hat{a}_R + \hat{a}^{\dagger}_{R}\hat{a}_L \right) -  U \hat{a}^{\dagger}_{L} \hat{a}_{L}\hat{a}^{\dagger}_{R}\hat{a}_{R},
\end{equation}
where $J$ and $U$ are expressed as appropriate integrals of the wave functions $\psi_L (\ve{x})$, $\psi_R (\ve{x})$. 
In eq.~\eqref{hamiltoniantwomode} we choose the convention to have $U$ positive (negative) for $a_S$ positive 
(negative), i.e. for repulsive (attractive) interactions. The distance $d$ between the wells varies 
typically between $\approx 0.5 $ $\mu$m and $\approx 5 $ $\mu$m: if the double well potential 
is created by an optical potential $V_{\text{opt}}=V_{\text{0}} \cos^{2}{kx}$ with $k=2\pi/\lambda_{\text{opt}}$ 
(being $d=\lambda_{\text{opt}}/2$), then for $^{87}Rb$ with $\lambda_{\text{opt}} \sim 1 \mu$m one has that being in the two-mode regime requires $V_{\text{0}}/h \gtrsim 5$ kHz~\cite{cataliotti,Ank05}, while for $\lambda_{\text{opt}} \sim 10$ $\mu$m is sufficient to have $V_{\text{0}}/h \gtrsim 500$ Hz~\cite{Alb05}. If the barrier is created by a laser beam having the maximum 
at the center of the trap, the distance is $\gtrsim 2 \mu$m  in \cite{Val15} with $^{6}Li$ atoms 
the barrier was created by a laser beam at 532 nm, blue-detuned with respect to the main optical 
transition of lithium atoms and at the trap center, the beam was Gaussian-shaped,
with a $1/e^2$ beam waist of $2 \mu$m. In this case, in the BEC regime 
the two-model regime was reached for $V_0/\mu \gtrsim 1.5$ with the chemical potential 
$\mu \sim 100 \hbar \omega$, with $\omega=15$ Hz. In all cases, of course, by increasing the barrier 
energy $V_{\text{0}}$ the ratio $U/J$ increases (see \textit{e.g.}~\cite{Mor06,bloch2008many}; a study of the dependence 
of the ratio for different dimensionality of the system can be found in~\cite{Tro05}). We finally observe that 
in presence of a negative scattering length, the coefficient $U$ can be negative (and yet the two-mode model 
be valid): the quantum phase transition occurs at a finite value of $\left \lvert U \right \rvert$ for which 
a population imbalance between the two wells has been recently observed~\cite{trenkwalder2016quantum}.

\section{States}

In this section we introduce the states whose collapse (and decoherence) 
is studied in the following of the paper.

\subsection{Phase state}

The creation of atomic coherent states using Bose Einstein condensates 
is a fundamental result for the study of many-body physics with ultracold atoms. 
Considering our case of interest, the phase state has the following 
expression~\cite{pitaevskii2016bose}:
\begin{equation}
\label{phasestate}
\ket{\phi} = \frac{1}{\sqrt{N!2^N}} \left( \hat{a}^{\dagger}_{L} + e^{i\phi}\hat{a}^{\dagger}_{R} \right)^N \ket{0}.
\end{equation}
The phase coherence properties of the above state is expressed by the off-diagonal elements of the single particle density matrix defined as follows 
\begin{equation}
\label{singleparticledensity} 
\rho^{(1)} = \frac{1}{N} 
\begin{pmatrix}
\langle \hat{a}^{\dagger}_{L} \hat{a}_L \rangle & \langle \hat{a}^{\dagger}_{L} \hat{a}_R \rangle  \\
\langle \hat{a}^{\dagger}_{R} \hat{a}_L \rangle & \langle \hat{a}^{\dagger}_{R} \hat{a}_R \rangle
\end{pmatrix}
,
\end{equation}
where $\langle \cdot \rangle = \bra{\phi} \cdot \ket{\phi}$. Using eq.~\eqref{phasestate} one has
\begin{equation}
\label{phasecoherence}
\langle \hat{a}^{\dagger}_{L} \hat{a}_R \rangle = N \frac{e^{i\phi}}{2}.
\end{equation}

Coherence properties of the state~\eqref{phasestate} are detected through the presence of interference 
fringes in the momentum density of the gas \cite{pitaevskii2016bose}. 
As discussed in appendix~\ref{AppA}, 
the average of the momentum density for a generic state in the symmetric double well 
potential~\eqref{hamiltoniantwomode} is
\begin{equation}
\label{momdens}
\, \, \langle \hat{a}^{\dagger}(\ve{p})\hat{a}(\ve{p}) \rangle = N \left \lvert \psi_L (\ve{p}) \right \rvert^2 \left \{1 + \frac{2}{N}\mathbb{R}\text{e} \left(e^{-i\frac{2 a p_x}{\hbar}}\langle \hat{a}^{\dagger}_{L} \hat{a}_R \rangle \right) \right \}
\end{equation}
where $p_x$ is the x-component of the momentum $\ve{p}$. Using eq.~\eqref{phasecoherence} 
for the phase state~\eqref{phasestate} in the expression~\eqref{momdens}, we obtain
\begin{equation}
\label{momdens2}
\, \, \langle \hat{a}^{\dagger}(\ve{p})\hat{a}(\ve{p}) \rangle = N \left \lvert \psi_L (\ve{p}) \right \rvert^2 \left \{1 + \cos \left(\phi - \frac{2 a p_x}{\hbar} \right) \right \}.
\end{equation}
We observe that the phase coherence~\eqref{phasecoherence} is not, generally, 
a constant of motion of the Hamiltonian~\eqref{hamiltoniantwomode}, 
due to the presence of the interaction term. In fact, in the case of negligible tunneling, 
\textit{i.e.} with $J = 0$, the time evolution of the phase coherence~\eqref{phasecoherence} is 
(see the derivation in appendix~\ref{AppB}) 
\begin{equation}
\label{3.phasecoherencetime}
\begin{split}
&\langle \hat{a}^{\dagger}_{L} \hat{a}_R \rangle_t = \bra{\phi} e^{\frac{i}{\hbar} \hat{H} t}  \hat{a}^{\dagger}_{L} \hat{a}_R e^{-\frac{i}{\hbar} \hat{H} t}  \ket{\phi} \\
&= N \frac{e^{i\phi}}{2} \left[ \cos \left ( \frac{tU}{\hbar} \right)\right]^{N-1}\approx  N \frac{e^{i\phi}}{2} e^{-N\left(\frac{tU}{\hbar}\right)^2},
\end{split}
\end{equation}
where the approximation in the last line holds for small time $t \ll \hbar / \mid U \mid$ and 
large number of atoms $N \gg 1$ (see the study of the 
phase diffusion in \cite{lewenstein1996quantum,javanainen1997phase}). 
The phase coherence is periodically reestablished with period 
$T=\hbar/ \mid U \mid$~\cite{imamoḡlu1997inhibition,greiner2002collapse,faust2010analysis}. 

\subsection{Superposition of phase states}

Bose Josephson junctions are also promising systems to create macroscopically entangled states. 
The first state considered in this paper is the superposition of phase states as given in~\eqref{phasestate}. 
They can be dynamically created from a single phase state, with dynamics given by Hamiltonian~\eqref{hamiltoniantwomode} with $J = 0$. Indeed, after a time $t_2 = \hbar\pi /(2U)$, the initial phase state~\eqref{phasestate} evolves in
\begin{equation}
\label{supphase}
\ket{\phi_{t_2}} = \frac{1}{\sqrt{2}}\left(\ket{\phi} + e^{i\beta}\ket{\phi +\pi}\right),
\end{equation}
where $\beta$ is a fixed phase difference between the two phase states. 
The state~\eqref{supphase} does not show any single-particle coherence properties~\eqref{phasecoherence}, 
but it shows $N$-particles coherence. As shown in appendix~\ref{AppB} one finds
\begin{equation}
\label{ncorr1}
\! \! \bra{\phi} \hat{a}^{\dagger^k}_{L} \hat{a}_{R}^{k} \ket{\phi+\pi}= 
\begin{cases}
0,  & \text{ if $k<N$;} \\
\frac{N!e^{i N \phi}}{2^{N}}  ,   & \text{ if $k=N$.} 
\end{cases}
\end{equation}
Eq.~\eqref{ncorr1} means that the way to distinguish the 
state~\eqref{supphase} from the statistical mixture 
$\hat{\rho} = \left(\ket{\phi}\bra{\phi} + 
\ket{\phi +\pi}\bra{\phi +\pi} \right)/2$ is through a measure of 
$N$-particles observables, \textit{i.e.}, the reduced density matrix at 
$k$-particles does not show any coherences, except if $k=N$. 

\subsection{NOON states}

The second macroscopically entangled state considered in this paper is the so-called NOON state, defined as
\begin{equation}
\label{noon}
\ket{\text{NOON}}= \frac{1}{\sqrt{2 N!}}\left(\hat{a}_{L}^{\dagger^N}+\hat{a}_{R}^{\dagger^N}\right) \ket{0}.
\end{equation}
The NOON state~\eqref{noon} is the ground state of the Hamiltonian~\eqref{hamiltoniantwomode} with $J = 0$ 
and $U<0$. Even though several proposals have been formulated to create a NOON state in a double well 
(see e.g.~\cite{Cir98,Dal00,Dun01,louis2001macroscopic,Maz07,Fer08}), its very short life-time 
with respect to decoherence makes its experimental realization an open problem. A first step in this direction has been done recently~\cite{trenkwalder2016quantum}, where the experimental realization of double well bosonic systems with negative and controllable scattering length was achieved.

A relation similar to eq.~\eqref{ncorr1} is found (see appendix~\ref{AppB})
\begin{equation}
\label{ncorr2}
\bra{\text{NOON}} \hat{a}^{\dagger^k}_{L} \hat{a}_{R}^{k} \ket{\text{NOON}}= 
\begin{cases}
0,  & \text{ if $k<N$;} \\
\frac{\left(N!\right)^2}{2},   & \text{ if $k=N$.} 
\end{cases}
\end{equation}
The meaning of eq.~\eqref{ncorr2} is the same as for eq.~\eqref{ncorr1}: the only way 
to distinguish the NOON state~\eqref{noon} from the statistical mixture 
$\hat{\rho}= \left(\ket{N_{L},0_{R}}\bra{N_{L},0_{R}} + \ket{0_{L},N_{R}}\bra{0_{L},N_{R}}\right)/2$ 
is by looking at $N$-particles observables. 

The fact that NOON state~\eqref{noon} and the superposition of coherent states~\eqref{phasestate} show similar N-particles coherences can be understood considering that the superposition of coherent states~\eqref{phasestate} can be obtained from a NOON state passing through a 50:50 beam splitter~\cite{gordon1999creating,Dun01}. The unitary single particle dynamics of the beam splitter does not destroy the initial N-particles coherence of the NOON state. As a consequence, the CSL effects on NOON state and superposition of phase states are the same, as we will see in the next section.

From this point of view, the difference between the atomic coherent state~\eqref{phasestate} and 
the macroscopic entangled states~\eqref{supphase} and~\eqref{noon} is quite evident. 
In fact, in the coherent state all the atoms are in the same superposition of single particle states. 
Single particle observables are enough to detect the quantumness of a coherent state. On the contrary, 
in the macroscopic entangled states~\eqref{supphase} and~\eqref{noon} 
the superposition is at the level of the whole system and, in order to detect it, the proper 
$N$-particles observables are required. Notice that differently from the 
coherence properties~\eqref{phasecoherence} of the phase state~\eqref{phasestate}, 
the $N$-particles coherences~\eqref{ncorr1} and~\eqref{ncorr2} are left unchanged 
by the Hamiltonian~\eqref{hamiltoniantwomode} with $J = 0$.

\section{CSL model for a Bose Josephson junction}\label{sec:CSL}

In the CSL model, the density matrix evolution for a bosonic system with $N$ particles is described by the following master equation~\cite{laloe2014heating}:
\begin{equation}
\label{csl}
\begin{split}
\frac{d\hat{\rho}(t)}{dt} = &-\frac{i}{\hbar} \left [\hat{H},\hat{\rho} (t) \right]- \frac{\lambda A^2}{2} \int d\ve{y} \int d\ve{y}' \, e^{-\frac{(\ve{y}-\ve{y}')^2}{4 r^{2}_{c}}} \times \\
&\left [\hat{a}^{\dagger} (\ve{y})\hat{a}(\ve{y}), \left[\hat{a}^{\dagger}(\ve{y}') \hat{a} (\ve{y}'), \hat{\rho} (t) \right ] \right],
\end{split}
\end{equation}
where $A$ is the number of nucleons in each atom, and $\lambda$ and $r_C$ 
are the parameters characterizing the model, as described in the introductive section.

In order to apply the CSL model to a Bose Josephson junction, 
it is convenient to rewrite the master equations~\eqref{csl} in terms of the left and right states. 
Using eq.~\eqref{creationtwomode} in eq.~\eqref{csl}, we get
\begin{equation}
\label{csltwomode}
\frac{d\hat{\rho}(t)}{dt} = -\frac{i}{\hbar} \left [\hat{H},\hat{\rho} (t) \right]- \frac{\lambda A^2}{2} \sum_{\substack{ i,j \\ k,l }} \gamma_{k,l}^{i,j} \left [\hat{a}^{\dagger}_{i}\hat{a}_j, \left[ \hat{a}^{\dagger}_{k}\hat{a}_l, \hat{\rho} (t) \right] \right],
\end{equation}
where
\begin{equation}
\label{gamma}
\gamma_{k,l}^{i,j} = \int d\ve{y} \int d\ve{y}' \, e^{- \frac{\abs{\ve{y} - \ve{y}'}^2}{4 r^{2}_{c}}}  \psi_{i}^{*} (\ve{y}) \psi_{j}(\ve{y}) \psi_{k}^{*} (\ve{y}') \psi_{l} (\ve{y}'),
\end{equation}
with $i,j,k,l = L,R$. We can immediately note that couplings with an odd number of the same index, 
like $\gamma_{L,L}^{L,R}$, are null, due to the orthogonality of the two modes. Therefore, 
we have to evaluate two types of couplings: {\em (i)} 
$\gamma_{L,L}^{R,R} = \gamma_{R,R}^{L,L} \equiv \gamma_{i,j\neq i}$, 
where the equality holds because of the parity in the CSL Gaussian term; 
and {\em (ii)} $\gamma_{L,L}^{L,L}=\gamma_{R,R}^{R,R} \equiv \gamma_{i,i}$, 
where the equality holds because of the symmetry of the double well potential. 

By taking into account that the total number of particle is fixed, \textit{i.e.}, 
$\hat{a}^{\dagger}_{L}\hat{a}_L + \hat{a}^{\dagger}_{R}\hat{a}_R = N$, 
it is possible to show that the dissipative term in the master equation~\eqref{csltwomode} 
takes the following expression:
\begin{equation}
\label{csldiss}
\begin{split}
&\sum_{\substack{ i,j= L,R}}\gamma_{i,j} \left [\hat{a}^{\dagger}_{i}\hat{a}_i, \left[ \hat{a}^{\dagger}_{j}\hat{a}_j, \hat{\rho} (t) \right] \right] \\
&= \sum_{i=j} \gamma_{i,i} \left [\hat{a}^{\dagger}_{i}\hat{a}_i, \left[ \hat{a}^{\dagger}_{i}\hat{a}_i, \hat{\rho} (t) \right] \right] +  \sum_{i \neq j}\gamma_{i,j} \left [\hat{a}^{\dagger}_{i}\hat{a}_i, \left[ \hat{a}^{\dagger}_{j}\hat{a}_j, \hat{\rho} (t) \right] \right] \\
& = \sum_i \left( \gamma_{i,i} - \gamma_{i,j\neq i} \right)\left [\hat{a}^{\dagger}_{i}\hat{a}_i, \left[ \hat{a}^{\dagger}_{i}\hat{a}_i, \hat{\rho} (t) \right] \right] \\
& = \bar{\gamma}  \sum_i \left [\hat{a}^{\dagger}_{i}\hat{a}_i, \left[ \hat{a}^{\dagger}_{i}\hat{a}_i, \hat{\rho} (t) \right] \right],
\end{split}
\end{equation}
where we have defined
\begin{equation}
\label{gammabarra}
\begin{split}
&\bar{\gamma} =  \int d\ve{y} \int d\ve{y}' \, e^{- \frac{\abs{\ve{y} - \ve{y}'}^2}{4 r^{2}_{c}}} \abs{\psi_{L} (\ve{y})}^2 \left(\abs{\psi_{L} (\ve{y}')}^2 - \abs{\psi_{R} (\ve{y}')}^2 \right) \\
& \approx 1-e^{-\frac{d^2}{4r_{C}^{2}}} \, ;
\end{split}
\end{equation}
the last equality is expected to be satisfied in the two-mode approximation.

The master equation~\eqref{csltwomode} then becomes
\begin{equation}
\label{cslbjj}
\frac{d\hat{\rho}(t)}{dt} = -\frac{i}{\hbar} \left [\hat{H},\hat{\rho} (t) \right]- \frac{\lambda A^2 \bar{\gamma}}{2} \sum_{\substack{ i= L,R}} \left [\hat{a}^{\dagger}_{i}\hat{a}_i, \left[ \hat{a}^{\dagger}_{i}\hat{a}_i, \hat{\rho} (t) \right] \right].
\end{equation}
To have an explicit solution, we neglect the hopping term of~\eqref{hamiltoniantwomode}. As shown in appendix~\ref{AppC}, the solution of eq.~\eqref{cslbjj} is
\begin{equation}
\label{cslmatrix}
\begin{split}
& \hat{\rho}(t) =  e^{ \frac{4iUt}{\hbar}\hat{a}^{\dagger}_{L}\hat{a}_{L}\hat{a}^{\dagger}_{R}\hat{a}_{R}}e^{- \lambda A^2 \bar{\gamma}t\left(\overset{\longleftarrow}{\hat{a}^{\dagger}_{R}\hat{a}_{R}}-\overset{\longrightarrow}{\hat{a}^{\dagger}_{R}\hat{a}_{R}}\right)^2} \hat{\rho}(0)  e^{-\frac{4iUt}{\hbar}\hat{a}^{\dagger}_{L}\hat{a}_{L}\hat{a}^{\dagger}_{R}\hat{a}_{R}} \\
&= e^{- \lambda A^2 \bar{\gamma}t\left(\overset{\longleftarrow}{\hat{a}^{\dagger}_{R}\hat{a}_{R}}-\overset{\longrightarrow}{\hat{a}^{\dagger}_{R}\hat{a}_{R}}\right)^2}\hat{\rho}_{\text{Sch}}(t),
\end{split}
\end{equation}
where $\overset{\longleftarrow}{\hat{a}^{\dagger}_{R}\hat{a}_{R}} \left(\overset{\longrightarrow}{\hat{a}^{\dagger}_{R}\hat{a}_{R}} \right)$ acts on the left (right) of the density matrix, and 
\begin{equation}
\label{4.densitymatrixsch}
\hat{\rho}_{\text{Sch}}(t) =  e^{ \frac{iUt}{\hbar}\hat{a}^{\dagger}_{L}\hat{a}_{L}\hat{a}^{\dagger}_{R}\hat{a}_{R}} \hat{\rho}(0)  e^{-\frac{iUt}{\hbar}\hat{a}^{\dagger}_{L}\hat{a}_{L}\hat{a}^{\dagger}_{R}\hat{a}_{R}}
\end{equation}
is the density matrix evolved under the Schr\"odinger (unitary) dynamics only.

\subsection{Collapse dynamics due to CSL}

Let us see how an initial phase state evolves under the CSL dynamics, looking in particular to the coherence 
properties given by eq.~\eqref{phasecoherence}. From eq.~\eqref{cslmatrix}, it is possible to see that 
(see appendix~\ref{AppC}):
\begin{equation}
\label{CSLphase}
\langle \hat{a}^{\dagger}_{L} \hat{a}_R \rangle^{\text{CSL}}_{t} = \text{Tr} \left[ \hat{a}^{\dagger}_{L} \hat{a}_R \hat{\rho} (t) \right] 
= e^{-\lambda A^2 \bar{\gamma}t} \langle  \hat{a}^{\dagger}_{L} \hat{a}_R \rangle^{\text{Sch}}_{t}.
\end{equation}
Taking into account the  Schr\"odinger evolution of the phase state coherence~\eqref{3.phasecoherencetime}, 
we obtain 
\begin{equation}
\label{CSLphase2}
\begin{split}
&\langle \hat{a}^{\dagger}_{L} \hat{a}_R \rangle^{\text{CSL}}_{t} = N e^{-\lambda A^2 \bar{\gamma}t} \frac{e^{i\phi}}{2} 
\left[ \cos \left ( \frac{tU}{\hbar} \right)\right]^{N-1} \\
& \approx  N \frac{e^{i\phi}}{2} e^{-\lambda A^2 \bar{\gamma}t} e^{-N\left(\frac{tU}{\hbar}\right)^2},
\end{split}
\end{equation}
where the last line is valid only for $t \ll \hbar U^{-1}$. 

From eq.~\eqref{CSLphase2}, it is clear that the action of the CSL dynamics is to break the spatial 
superposition~\eqref{phasecoherence} of the phase state. Since, in this case, 
the many-body state is a coherent factorization of delocalized single particle states, 
then an amplification mechanism is missing, and the CSL collapse rate does not depend on the 
total number of particles $N$. Thus, the use of phase states such as~\eqref{phasestate} 
to set experimental bounds on the CSL parameters is not convenient in general. 

\begin{figure}[t]
\includegraphics[scale=0.43]{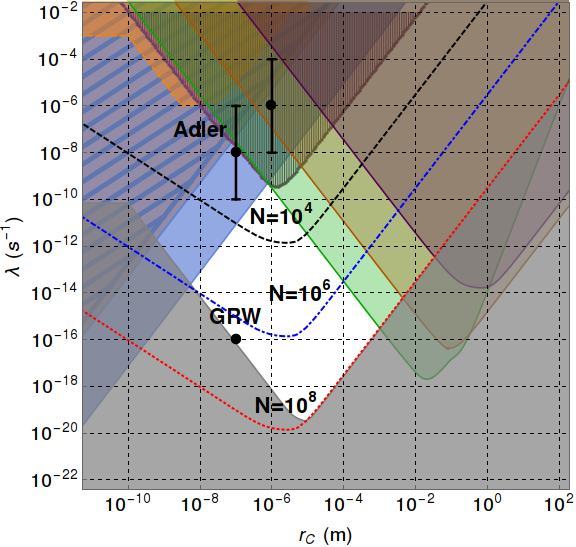}
\caption{Most recent exclusion plot of the CSL
model~\cite{carlesso2016experimental}.  The upper coloured regions represent the values of $\lambda$ and $r_C$ that have been excluded by  experimental data. The lower grey region is a lower bound coming from the requirement that CSL must be effective in localising macroscopic objects. The white region---that of interest---is still unexplored.  The picture shows
 the bounds of an hypothetical experiment involving the
macroscopically entangled states~\eqref{supphase} or \eqref{noon}. Here,
we made the hypothesis that their $N$-particles coherences are preserved
for a time $t=1$ s, with three different number of particles: $N=10^4$
(black dashed line), $10^6$ (blue dot-dashed line) and $10^8$ (red dotted
line). The excluded regions are the upper part of the related lines.}
\label{excplot}
\end{figure}

As for example, let us consider the experiment~\cite{jo2007long}, where the long 
coherence time of $200$ ms with a gas of $^{23}$Na atoms was observed. Using eq.~\eqref{gammabarra}, 
we have that the exponential decrease of the coherence in eq.~\eqref{CSLphase2} induced by CSL dynamics 
becomes experimentally visible only if $\lambda \geq 1/(t A^2)$. Then from eq.~\eqref{CSLphase2} 
it is possible to put a bound of $\lambda \leq 10^{-2}$ s$^{-1}$. This is a weak bound for two main reasons: 
firstly, the expected values of the collapse rate $\lambda$ theoretically predicts by Ghirardi, Rimini and 
Weber ($\lambda = 10^{-16}$  s$^{-1}$) and by Adler 
($\lambda = 10^{-8 \pm 2}$  s$^{-1}$ and $\lambda = 10^{-6 \pm 2}$  s$^{-1}$) 
are much smaller than $10^{-2}$ s$^{-1}$; secondly, much stronger bounds have been set by 
other experiments, as can be seen in the exclusion plot~\ref{excplot}.
\begin{figure}[t]
\includegraphics[scale=0.43]{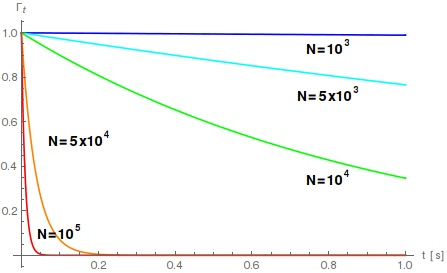}
\caption{Time evolution~\eqref{CSLncorr} of the normalized $N$-particles coherences $\Gamma_t = \langle \hat{a}^{\dagger}_{L} \hat{a}_R \rangle^{\text{CSL}}_{t}/ \langle \hat{a}^{\dagger}_{L} \hat{a}_R \rangle^{\text{CSL}}_{0}$, for different number of atoms. Here we fixed the CSL
parameters $\lambda = 10^{-11} \, \text{s}^{-1}$ and $r_C = 10^{-7}$
m, which are among the weakest parameters of the white region in fig.~\ref{excplot}.} 
\label{plottime}
\end{figure}
Let us now consider the CSL evolution of the macroscopically entangled 
states~\eqref{supphase} and~\eqref{noon}. In this case, the $N$-particles coherence 
properties~\eqref{ncorr1} and~\eqref{ncorr2} are preserved by 
the Schr\"odinger evolution. Using eq.~\eqref{cslmatrix}, as discussed 
in appendix~\ref{AppC} it is possible to see that 
\begin{equation}
\label{CSLncorr}
\langle \hat{a}^{\dagger^N}_{L} \hat{a}_{R}^{N} \rangle^{\text{CSL}}_{t} = e^{-\lambda N^2 A^2 \bar{\gamma}t} \langle  \hat{a}^{\dagger^N}_{L} \hat{a}_{R}^{N} \rangle^{\text{Sch}}_{t}.
\end{equation}
Differently from the phase decoherence case~\eqref{CSLphase2}, the exponential decoherence 
of the macroscopically entangled states~\eqref{CSLncorr} is faster due to a factor $N^2$. 
This is due to the macroscopicity of the states~\eqref{supphase} and~\eqref{noon},\textit{i.e.}, 
these are superposition of localized $N$-particles states, which is expressed 
by the $N$-particles quantum coherences~\eqref{ncorr1} and~\eqref{ncorr2}. 
As a consequence, an amplification mechanism occurs in the CSL dynamics, increasing the collapse rate 
by a factor $N^2$.

 In fig.~\ref{excplot} we show the exclusion plot in the $\lambda - r_C$ parameters space of CSL~\cite{carlesso2016experimental}. The upper coloured regions indicate the excluded values of the parameters set by experimental data. The lower grey region comes from requiring that the CSL mechanism is strong enough to localize macroscopic objects~\cite{torovs2016bounds}, that is the main motivation that introduced collapse models in first place~\cite{bassi2003dynamical}. Fig.~\ref{excplot}  shows also the exclusion plot for an hypothetical experiment 
involving the macroscopically entangled states~\eqref{supphase} or~\eqref{noon}. Here, we had to fix 
a time $t$ for which the $N$-particles coherences~\eqref{ncorr1} and~\eqref{ncorr2} 
are preserved and observed. The longer $t$, the smaller the number of atoms 
needed to use (the dependence on the time is linear, the one on the number is quadratic). An inspection 
of results show that in order to have both reasonable values for the particle number $N$ and at the same time 
explore the white region in fig.~\ref{excplot}, the one not yet excluded by experiments, 
one has to use $t \sim 1$ s. This is presently a very challenging request, and the goal of the 
present computation is to clarify and predict what value of $t$ is needed to explore the uncovered region of the CSL parameter space for $N$ going from $10^2$ to $10^6$.

We then choose $t=1$ s and we considered a $Rb$ gas, with width of each well given by $\sigma$, 
and with the two wells distant $d$. With $d=10\mu$m $J$ is negligible, and the results 
for $J=0$ apply and are plotted in fig.~\ref{excplot} for three different values 
of $N$ (with $\sigma \approx 1 \mu$m). Also reducing $d$ in order to have a finite $J$ does not quantitatively 
change the region that can be explored in the $\lambda-r_C$ space (actually, we expect it shrinks) - similarly 
increasing $\sigma$, which results in decreasing $U$, does not change the boundaries of the region, as soon 
as the two-mode approximation holds. We conclude that with $t=1$ s one needs $N \sim 10^3-10^4$ to 
enter the white region not yet explored and that with the (presentely prohibitive) number $N=10^8$ the whole white region 
can be probed. From fig.~\eqref{plottime}, where the time evolution~\eqref{CSLncorr} is represented,  
the strong dependence on $N$ is clear.

All the computations above have been done under the hypothesis that the unitary Schr\"odinger dynamics is modified only by CSL noise, without taking into account other decoherence sources. In the next section we compare the main decoherence sources with CSL, and we give estimates on the conditions that experiments must fulfill in order to properly test CSL model.

\section{Environment decoherence on Bose Josephson junctions and comparison with the CSL model}
\label{deco}

Problems in testing CSL dynamics in experiments usually arise from environmental decoherence sources, 
leading to similar loss of coherences. This is the topic of this section, in which 
we compare the decoherence of the CSL noise with the decoherence induced by the external environment. 
In particular, we focus on four main sources, i.e., the thermal cloud surrounding the condensate, 
the three body recombination processes, phase noise 
and the trapping laser. We then study the conditions that 
experiments have to fulfil in order to reduce  environment decoherence, which is a necessary condition 
to detect CSL effects. For each decoherence source we consider realistic values of the parameters from typical experimental setups, see e.g.~\cite{burt1997coherence,laloe2014heating,trenkwalder2016quantum}. It turns out that these  setups generally present environmental decoherence that covers any CSL effect within the white region of fig.~\ref{excplot}. We then estimate what are the values of experimental setup parameters needed to lower environmental decoherence and test CSL according to the analysis described in the previous section.

\subsection{Interaction with a thermal cloud}

Even if the temperature is very low, a small amount of thermally excited atoms is always present. 
Atoms in excited states interact with the condensate, leading to two main effects: 
a loss of atoms from the condensate to the thermal cloud; and a decoherence on the condensate in the position basis. 
As reviewed in appendix~\ref{AppD}, 
the master equation describing the dynamics of a condensate interacting with its thermal cloud is:
\begin{equation}
\label{therm1}
\frac{d\hat{\rho}(t)}{dt} = -\frac{i}{\hbar} \left [\hat{H},\hat{\rho} (t) \right] + \sum_{i= 1,2}\mathcal{L}_{t}^{(i)} \left[ \hat{\rho}(t) \right] \, ,
\end{equation}
where $\mathcal{L}_{t}^{(1)} \left[ \hat{\rho}(t) \right]$ describes the atom losses dynamics, 
which can be written in the following way for the particular case of a condensate in a double well 
potential~\cite{sinatra1998phase}:
\begin{equation}
\label{loss}
\mathcal{L}_{t}^{(1)} \left[ \hat{\rho}(t) \right] = \Lambda_{\text{loss}}\sum_{j = L, R} \, \hat{a}_j \hat{\rho} (t) \hat{a}_{j}^{\dagger} - \frac{1}{2} \left \{\hat{a}_{j}^{\dagger} \hat{a}_{j}, \hat{\rho} (t) \right \} \, ,
\end{equation}
where $\Lambda_{\text{loss}}$ is the rate of atom losses from the condensate to the thermal cloud, which, 
according the discussion in Appendix~\ref{AppD} is given by the following local correlation function of the thermal cloud 
\begin{equation}
\label{lamlos}
\begin{split}
&\Lambda_{\text{loss}} = \frac{g^2}{\hbar^2} \int d\ve{x} \, \abs{\psi_{L} (\ve{x})}^2 \times\\
&\times \text{Tr} \left [\hat{a}^{\dagger} (\ve{x})\hat{a}^{\dagger} (\ve{x})\hat{a}^{\dagger} (\ve{x})\hat{a} (\ve{x})\hat{a} (\ve{x})\hat{a} (\ve{x}) \hat{\rho}_{\text{therm}}\right] \, .
\end{split}
\end{equation}
The term $\mathcal{L}_{t}^{(2)} \left[ \hat{\rho}(t) \right]$ introduced in eq.~\eqref{therm1} 
describes interactions between the condensate (C) and non condensate (NC) atoms of the type $C + NC \to C + NC$. 
This interaction is in position, and leads to a decoherence dynamics that preserves the populations of the gas. 
In particular, for a double well potential, the following identity holds
\begin{equation}
\label{cnc}
\mathcal{L}_{t}^{(2)} \left[ \hat{\rho}(t) \right] = - \frac{\Lambda_{\text{dec}}}{2}\sum_{j = L, R} \,  \left [\hat{a}^{\dagger}_{j}\hat{a}_j, \left[ \hat{a}^{\dagger}_{j}\hat{a}_j, \hat{\rho} (t) \right] \right] \, ,
\end{equation}
where
\begin{equation}
\label{lamdec}
\begin{split}
&\Lambda_{\text{dec}} = \frac{g^2}{\hbar^2} \int d\ve{x} \, \text{Tr} \left [\hat{a}^{\dagger} (\ve{x})\hat{a} (\ve{x})\hat{a}^{\dagger} (\ve{x})\hat{a} (\ve{x}) \hat{\rho}_{\text{therm}}\right] \times \\
&\times \abs{\psi_{L} (\ve{x})}^4 \, .
\end{split}
\end{equation}
Let us consider first the term in eq.~\eqref{lamlos}. Neglecting the Hamiltonian dynamics, 
it is easy to see that the $N$-particles correlation has the following time dependence:
\begin{equation}
\label{lossncorr}
\langle \hat{a}^{\dagger^N}_{L} \hat{a}_{R}^{N} \rangle^{\text{loss}}_{t} = e^{-\Lambda_{\text{loss}} N t} \langle  \hat{a}^{\dagger^N}_{L} \hat{a}_{R}^{N}\rangle_{0}.
\end{equation}
Differently from the CSL case~\eqref{CSLncorr}, the decoherence rate induced by atom losses 
increases linearly with the number of atoms. Therefore, in order to test CSL effects on a NOON state, we need CSL decoherence~\eqref{CSLncorr} to be stronger than the decoherence induced by atom losses~\eqref{lossncorr}. This means that, if we want to experimentally test CSL in the white region of fig.~\ref{excplot}, then we must have $N \lambda A^2 \bar{\gamma} \geq \Lambda_{\text{loss}}$, with $\bar{\gamma}$ given by eq.~\eqref{gammabarra}. In the particular case of $r_{C} = 10^{-7}$ m~\cite{ghirardi1990markov}, $\bar{\gamma} \approx 1$ for typical distances $d\gtrsim 0.5 \mu$m between the wells (see the discussion in Sec.~\ref{sec:bjj}).

The one give here above is the general condition to test CSL effects against atom losses. Let us now consider, for example, the experimental setup described 
in~\cite{burt1997coherence}, where the authors measured $\Lambda_{\text{loss}} \approx 4 \times 10^{-3}$ s$^{-1}$ 
for a non-homogeneous gas at fractional temperature $T/T_C \approx 1/3$, whit $T_C$ critical temperature 
of the gas. Comparing the decay rates of the CSL model in eq.~\eqref{CSLncorr} with the decay given 
by the atomic losses as in eq.~\eqref{lossncorr} with the parameters given in~\cite{burt1997coherence}, 
the CSL damping turns out to be faster only if the NOON state  is composed by 
$N > \Lambda_{\text{loss}} / \left(\lambda A^2 \right) \approx 10^{10}$ atoms, considering the value 
$\lambda = 10^{-17}$ s$^{-1}$ as proposed in~\cite{ghirardi1990markov}, and $A \approx 100$. 
From eq.~\eqref{lamlos}, the decoherence rate induced by the atomic losses can be reduced either by a proper 
decreasing of the temperature of the gas (so reducing the thermal density) or by decreasing the coupling 
interaction, which can be achieved by use of Feshbach resonances. We have that
\begin{equation}
\label{lamlos2}
\Lambda_{\text{loss}} \approx \frac{g^2 n_{\text{therm}}^{3}}{\hbar^2}.
\end{equation}
In order to detect CSL effects on a NOON state with, for example $N=10^8$ atoms 
(the number of atoms we need in order to probe the whole unexplored region of the parameter space, 
as shown in fig.~\ref{excplot}), eq.~\eqref{lamlos2} shows that the experimental setup described 
in~\cite{burt1997coherence} must be modified either by reducing the thermal density or by decreasing 
the coupling constant of one order of magnitude.

Let us now focus on the term in eq.~\eqref{cnc}. We immediately note the similarities between 
eq.~\eqref{cnc} and the CSL master equation~\eqref{cslbjj}, with $\lambda A^2 \to \Lambda_{\text{dec}}$. 
Through eqs.~\eqref{lamdec} and~\eqref{lamlos2}, it is possible to relate the decoherence rate due 
to the atomic losses to the decoherence rate described in eq.~\eqref{cnc} as follows:
\begin{equation}
\label{lamdec2}
\Lambda_{\text{dec}} \approx \frac{\Lambda_{\text{loss}}}{N_{\text{therm}}} \, ,
\end{equation}
where $N_{\text{therm}}$ is the number of atoms in the thermal cloud. As before, we can experimentally test CSL within the white region of fig.~\ref{excplot} if $\lambda A^2 \bar{\gamma} \geq \Lambda_{\text{dec}}$.

As an example, let us consider again the experiment described in~\cite{burt1997coherence}, with $N_{\text{therm}} \approx N \left(T/T_C \right)^3 \approx 10^4$, where we used the relation between the 
total number of atoms with the number in the thermal cloud of an ideal gas in an harmonic trap~\cite{pitaevskii2016bose}. Using then eq.~\eqref{lamdec2} we have $\Lambda_{\text{dec}} \approx 4 \times 10^{-10}$ s$^{-1}$. If $\lambda = 10^{-17}$ $\text{s}^{-1}$ and $r_{C} = 10^{-7}$ m, then $\bar{\gamma} \approx 1$, and $\Lambda_{\text{dec}} < \lambda A^2 \approx 10^{-13}$ s$^{-1}$. Accordingly, the experimental setup described in~\cite{burt1997coherence} must be modified either by reducing the thermal density or by decreasing the coupling constant by two orders of magnitude.

\subsection{Decoherence induced by three body recombination processes}

We consider in this section 
the effect of three body recombination processes, according to which atoms leave the trap, 
with negligible probability that they interact again with other atoms in the trap. 
We write the effective master equation describing the dynamics of a Bose Einstein condensate with three body 
recombination processes~\cite{Kag85,fedichev1996three,sinatra1998phase,jack2002decoherence} in the form
\begin{equation}
\label{threebody1}
\frac{d\hat{\rho}(t)}{dt} = -\frac{i}{\hbar} \left [\hat{H},\hat{\rho} (t) \right] + \Lambda_3 \sum_{j = L, R} \, \hat{a}_{j}^3 \hat{\rho} (t) \hat{a}_{j}^{\dagger^3} - \frac{1}{2} \left \{\hat{a}_{j}^{\dagger^3} \hat{a}_{j}^{3}, \hat{\rho} (t) \right \} \, ,
\end{equation}
where
\begin{equation}
\label{lam3}
\Lambda_{3} \approx \frac{\hbar a_{S}^{4}}{m} \int d\ve{x} \left \lvert \psi_L \left( \ve{x}\right)\right \rvert^6 \, .
\end{equation}
Using eqs.~\eqref{threebody1} and~\eqref{lam3}, one finds the following relation:
\begin{equation}
\label{threebody2}
\langle \hat{a}^{\dagger^N}_{L} \hat{a}_{R}^{N} \rangle_{3}\left( t\right) = e^{- \frac{\hbar a_{S}^{4} n_{\text{BEC}}^{2}}{m} N t} \langle  \hat{a}^{\dagger^N}_{L} \hat{a}_{R}^{N} \rangle_{0} \, ,
\end{equation}
where $n_{\text{BEC}}$ is the condensate density. Comparing the three body 
decoherence rate in eq.~\eqref{threebody2} with the CSL decoherence rate $\lambda A^2 N^2$, we have a 
lower bound on the number of atoms of the NOON state:
\begin{equation}
\label{lam3_bis}
N \geq \frac{\hbar a_{S}^{4}n_{\text{BEC}}^{2}}{m \lambda A^2}.
\end{equation}
Consider, for example, the experimental measure described in~\cite{burt1997coherence}, 
where $\hbar a_{S}^{4}/ m \approx 5\times10^{-30}$ $\text{cm}^6/\text{s}$ and the peak density of the condensate 
is $5\times 10^{14}$ $\text{cm}^{-3}$. In this case, the CSL decoherence is faster if $N \geq 10^{13}$ atoms, 
for the particular case of  $\lambda = 10^{-17}$ s$^{-1}$. The three body decoherence 
rate can be decreased either by reducing the condensate density or by reducing the scattering length through 
the use of Feshbach resonance. For example, if we consider an experiment with a condensate density 
of $\approx10^{13}$ $\text{cm}^{-3}$ and a scattering length $a_S \leq 10^{-9}$ m, the CSL localization dynamics 
is dominant with $N \geq 10^7$ atoms. 

\subsection{Phase noise}

A very common technique to trap atoms is by using an external laser source. Thanks to the large 
experimental control on the laser electric field, several external potentials can be realized, 
with a large control on the parameters of the trap~\cite{metcalf2007laser}. However, every
optical trap is also a decoherence source for the trapped atomic system, leading to decoherence 
of the atomic density matrix in the position basis. Two of the main decoherence 
sources due to the laser are a phase noise, induced by fluctuations of the laser beam 
pointing~\cite{PhysRevA.84.043628}, and spontaneous photon emission processes from the 
atoms~\cite{cohen1977frontiers,cohen1990atomic,pichler2010nonequilibrium}. 

The phase noise in a Bose Josephson junction can be studied  
and treated as a stochastic noise modifying the energy levels of the system~\cite{PhysRevA.84.043628}. 
The effective density matrix evolution is given by:
\begin{equation}
\label{matrixnoise}
\hat{\rho} (t) = \int_{-\infty}^{+\infty} d\phi \, f(\phi,t) e^{-i\phi \hat{a}_{R}^{\dagger}\hat{a}_{R}}\hat{\rho}_{\text{Sch}}(t)  e^{i\phi \hat{a}_{R}^{\dagger}\hat{a}_{R}}.
\end{equation}
If the phase noise is a Gaussian noise with null average, then we have that
\begin{equation}
\label{f}
f(\phi,t) = \frac{1}{\sqrt{2 \pi} \Delta(t)} e^{-\frac{\phi^2}{2 \Delta^2 (t)}}.
\end{equation}
where the variance $\Delta^2 (t)$ completely characterizes the noise. Using eq.~\eqref{f} in 
eq.~\eqref{matrixnoise}, the density matrix evolution of a Bose Josephson junction under a Gaussian phase noise is:
\begin{equation}
\label{matrixnoise2}
\hat{\rho}(t) = e^{- \frac{\Delta^2 (t)}{2}\left(\overset{\longleftarrow}{\hat{a}^{\dagger}_{R}\hat{a}_{R}}-\overset{\longrightarrow}{\hat{a}^{\dagger}_{R}\hat{a}_{R}}\right)^2}\hat{\rho}_{\text{Sch}}(t).
\end{equation}
By comparing eq.~\eqref{matrixnoise2} with the density matrix 
evolution under the CSL dynamics eq.~\eqref{cslmatrix}, it is clear that the latter one can be tested only if the phase noise is reduced enough in the experiment. 
In~\cite{laloe2014heating}, the authors studied the effect of the fluctuations of the laser beam 
pointing on a single well potential for a gas of $^{133}$Cs atoms. In this work, the heating 
effect induced by the phase noise turns out to be negligible, compared to
the heating of the CSL model, if $\lambda > 10^5 (r_C/(1 \, \text{m}))^2$ s$^{-1}$. 
If we choose the value $r_C = 10^{-7}$ m (proposed by Ghirardi et al.l~\cite{ghirardi1986unified,ghirardi1990markov}), 
then, for the experimental setup described in~\cite{laloe2014heating}, the phase noise becomes negligible 
if $\lambda > 10^{-9}$ s$^{-1}$. This means that the interesting part of the parametric space shown in 
fig.~\ref{excplot} cannot be explored by such experimental setup due to a too strong phase noise. For 
example, in order to test CSL effects with the parameters 
$\lambda = 10^{-16}$ s$^{-1}$ $r_C = 10^{-7}$ m, then the phase noise should be reduced by a power 
$10^{-7}$, showing the very high control on the phase noise needed to explore the $\lambda-r_c$  parameter space.

\subsection{Spontaneous photon emission}

Another decoherence source related to optical traps is due to the spontaneous photon emission. 
In~\cite{pichler2010nonequilibrium}, the authors derived the master equation for two levels atoms in a 
far-detuned optical trap, which is given by:
\begin{equation}
\label{spont}
\begin{split}
\frac{d\hat{\rho}(t)}{dt} = &-\frac{i}{\hbar} \left [\hat{H},\hat{\rho} (t) \right]- \frac{\Gamma}{8 \delta^2} \int d\ve{y} \int d\ve{y}' \, \Omega (\ve{y}) \Omega (\ve{y}')\times \\
& \text{F} \left(k\left(\ve{y} -\ve{y}' \right) \right)\left [\hat{a}^{\dagger} (\ve{y})\hat{a}(\ve{y}), \left[\hat{a}^{\dagger}(\ve{y}') \hat{a} (\ve{y}'), \hat{\rho} (t) \right ] \right],
\end{split}
\end{equation}
where $\Gamma$ is the spontaneous emission rate; $\delta = \omega_l - \omega_0$, $\omega_l$ is the frequency of the 
laser, $\omega_0$ is the frequency resonance of the two levels atoms; $k = \omega_0/c$ with $c$ the 
speed of light in vacuum; $\Omega (\ve{x})$ Rabi frequency, related to the effective optical trap by
\begin{equation}
\label{opttrap}
V (\ve{x}) = \hbar \frac{\left \lvert \Omega (\ve{x})\right \rvert^2}{4 \delta}.
\end{equation}
The function $ \text{F} \left(\ve{z}\right)$ in eq.~\eqref{spont} is defined by the following integration:
\begin{equation}
\label{F}
\text{F} \left(\ve{z}\right) = \int_{\norma{\ve{u}} = 1} d\ve{u} \, e^{-i\ve{u} \cdot \ve{z}}.
\end{equation}
We note that the spontaneous emission master equation~\eqref{spont} has the same form as that of the CSL master
equation~\eqref{csl}, with the replacement:
\begin{equation}
\label{spontcsl}
\lambda A^2 e^{-\frac{(\ve{y}-\ve{y}')^2}{4 r^{2}_{c}}} \to  \frac{\Gamma}{4 \delta^2} \Omega (\ve{y}) \Omega (\ve{y}') \text{F} \left(k\left(\ve{y} -\ve{y}' \right) \right).
\end{equation}
Using eq.~\eqref{creationtwomode} in eq.~\eqref{spont}, with a similar procedure used to derive eqs.~\eqref{csltwomode} and~\eqref{cslmatrix}, it is possible to write down the density matrix for a gas of atoms trapped in an optical trap with spontaneous emission process:
\begin{equation}
\label{matrixspont}
\hat{\rho}(t) = e^{- \frac{\Gamma \bar{\Omega}t}{4 \delta^2}\left(\overset{\longleftarrow}{\hat{a}^{\dagger}_{R}\hat{a}_{R}}-\overset{\longrightarrow}{\hat{a}^{\dagger}_{R}\hat{a}_{R}}\right)^2}\hat{\rho}_{\text{Sch}}(t),
\end{equation}
where
\begin{equation}
\label{omegabarra}
\begin{split}
\bar{\Omega} = & \int d\ve{y} \int d\ve{y}' \, \Omega (\ve{y}) \Omega (\ve{y}') \text{F} \left(k\left(\ve{y} -\ve{y}' \right) \right) \\ 
&\abs{\psi_{L} (\ve{y})}^2 \left( \abs{\psi_{L} (\ve{y}')}^2 - \abs{\psi_{R} (\ve{y}')}^2\right).
\end{split}
\end{equation}
Compared to the CSL dynamics given in eq.~\eqref{cslmatrix}, the spontaneous emission processes 
are negligible only if $\lambda A^2 \bar{\gamma} \gg \Gamma \bar{\Omega}/\left( 4 \delta^2 \right)$. 
For example, let us consider the experimental setup described in~\cite{trenkwalder2016quantum}, 
where a laser with wavelength of $1064$ nm is used to trap a gas of $^{39}$K atoms. 
The resonance frequency of $^{39}$K is $\omega_0 \approx 390$ THz. 
Setting $r_C =10^{-7}$ m, we have that hypothetical CSL effects would be visible only 
if $\lambda > 10^{-12}$ s$^{-1}$. This means that the decoherence induced by spontaneous photon 
emission is strong enough to cover a large part of the parametric space shown in fig.~\ref{excplot}. 
In order to probe the interesting part of the CSL parametric space, the spontaneous photon emission in ~\cite{trenkwalder2016quantum} should be reduced by a power $10^4$. We conclude that the laser decoherence effects discussed in this and the previous sections are alleviated by using magnetic traps~\cite{RevModPhys.79.235}. Thermal and three-body effects cannot be fully removed. Thermal decoherence can be reduced either by weakening the atom-atom interaction, or by lowering the cloud density; three-body effects can be reduced again by weakening the interaction, or by reducing the condensate's density. This would allow to decrease the number of atom required to test CSL in the white region of fig.~\ref{excplot}, from the values discussed in this section to the values presented in Sec.~\ref{sec:CSL}.

\section{Conclusions}

Given the challenge to probe with current experiments the whole white region (not yet explored) in the CSL exclusion plot~\ref{excplot}, and motivated by the fact that nowadays experiments with Bose gases in double well potentials are able to create and detect strongly correlated many-body entangled states~\cite{Sch16}, we studied the possibility to test Continuous Spontaneous Localization (CSL) dynamics in a Bose Josephson junction of ultracold bosons. We also determined what requests the experimental values have to satisfy in order to put new bounds for the parameters of the CSL model in the white region of fig.~\ref{excplot}.

The collapse noise localizes the wave function in the position, which means that in a double well there is a localization in the left or in the right states. By solving the CSL exact master equation in the limit of negligible 
tunneling between the two wells of the potential, we found that the coherence of a phase state is slowly decreased 
by the localization process of the CSL model. At variance the CSL effects are much stronger in macroscopically entangled states 
as superposition of phase states and NOON states. 
Their $N$-atom coherences are suppressed by a factor 
exponentially depending on the number of atoms, 
leading to a fast localization process that can be used to test the CSL model. 
We also compared the CSL dynamics with two typical decoherence sources, 
namely, the phase noise and the spontaneous photon emission process. 
Their density matrix evolution mimics the CSL dynamics, and, usually, they are strong enough to cover CSL effects. We discussed under which conditions CSL effects would become stronger than these decoherence sources. We also concluded that magnetically trapped systems are more suitable to test the CSL model. We also compared the results of the CSL dynamics with thermal and three body effects.

Our analysis determines the bounds on the CSL parameters obtainable in experiments with cold atoms. The results are summarized in 
fig.~\ref{excplot}. In the $\lambda-r_C$ parameter space there is presently a region not yet explored, and we pointed out that having entangled states, as the NOON state, with $N \gtrsim 10^3$ and coherence time $t \gtrsim 1$ s would make possible 
to probe such a region. A similar outcome may be obtained when considering entangled states in an array of weakly coupled ultracold bosonic gases described by the Bose-Hubbard model~\cite{Bil16_t} or by varying the distance between (and the width of) the wells.
 
Although very demanding for current-day experiments, our results together with the recent very promising results on the implementation of squeezed states and the detection multi-particle correlations~\cite{Sch16} 
show that in perspective further advancements in the manipulation of highly-entangled states may open 
the possibility to study collapse models and to test quantum mechanics with ultracold atoms in double well potentials.

\appendix
\section{Momentum distribution}
\label{AppA}

From a Fourier transform of eq.~\eqref{creationtwomode} one finds
\begin{equation}
\label{twomodemom}
\hat{a} (\ve{p}) = \psi_L (\ve{p}) \hat{a}_L +\psi_R (\ve{p}) \hat{a}_R.
\end{equation}
For parity symmetry of the double well potential, we can relate the left and right states as follows:
\begin{equation}
\label{lrsym}
\psi_R (\ve{p}) = e^{-i\frac{d p_x}{\hbar}}\psi_L (\ve{p}),
\end{equation}
where we imposed that the right well is displaced $d$ from the left well along the x-direction.

Using eqs.~\eqref{twomodemom} and~\eqref{lrsym}, the momentum density operator, averaged over the phase state given in eq.~\eqref{phasestate}, becomes:
\begin{equation}
\label{momdens_bis}
\begin{split}
&\langle \hat{a}^{\dagger}(\ve{p})\hat{a}(\ve{p}) \rangle = \left \lvert \psi_L (\ve{p}) \right \rvert^2 \langle \hat{a}^{\dagger}_{L}\hat{a}_L \rangle + \left \lvert \psi_R (\ve{p}) \right \rvert^2 \langle \hat{a}^{\dagger}_{R}\hat{a}_R \rangle + \\
&2 \mathbb{R}\text{e} \left( \psi_{L}^{*} (\ve{p})\psi_R (\ve{p}) \langle \hat{a}^{\dagger}_{L} \hat{a}_R \rangle \right)\\
& = N \left \lvert \psi_L (\ve{p}) \right \rvert^2 \left \{1 + \cos \left(\phi - \frac{d p_x}{\hbar} \right) \right \},
\end{split}
\end{equation}
%
where we also imposed a fixed number $N$ of atoms.

\section{Time-dependent expectation values}
\label{AppB}

From the definition of phase state, using the Hamiltonian~\eqref{hamiltonian} with $J = 0$, we have
\begin{equation}
\label{phasecoherencetime}
\begin{split}
&\langle \hat{a}^{\dagger}_{L} \hat{a}_R \rangle_t = \bra{\phi} e^{\frac{i}{\hbar} \hat{H} t }  \hat{a}^{\dagger}_{L} \hat{a}_R e^{-\frac{i}{\hbar} \hat{H} t }  \ket{\phi} \\
& = \frac{1}{N! 2^N} \sum_{j,k = 0}^{N} \binom{N}{k} \binom{N}{j} e^{i\phi(j-k)} \bra{0} \hat{a}_{L}^{N-k}\hat{a}_{R}^{k} \\
&  e^{-\frac{i}{\hbar} t U \hat{a}^{\dagger}_{L} \hat{a}_{L}\hat{a}^{\dagger}_{R}\hat{a}_{R} } \hat{a}^{\dagger}_{L} \hat{a}_{R} e^{\frac{i}{\hbar} t U \hat{a}^{\dagger}_{L} \hat{a}_{L}\hat{a}^{\dagger}_{R}\hat{a}_{R} } \left(\hat{a}^{\dagger}_{L} \right)^{N-j} \left(\hat{a}^{\dagger}_{R} \right)^j \ket{0} \\
& = \frac{1}{N! 2^N} \sum_{j,k = 0}^{N} \binom{N}{k} \binom{N}{j} e^{i\phi(j-k)} e^{-\frac{i}{\hbar}tUk(N-k)} \times \\
& e^{\frac{i}{\hbar}tUj(N-j)}  \bra{0} \hat{a}_{L}^{N-k}\hat{a}_{R}^{k+1}\left(\hat{a}^{\dagger}_{L} \right)^{N-j+1} \left(\hat{a}^{\dagger}_{R} \right)^j \ket{0} \\
&=  \frac{e^{i\phi}}{N! 2^N} \sum_{k = 0}^{N-1} \binom{N}{k} \binom{N}{k+1} \times \\
& e^{\frac{i}{\hbar} tU [(k+1)(N-k-1) - k(N-k)]} (N-k)! (k+1)! \\
&= N \frac{e^{ i \left[\phi + (N-1) \frac{tU}{\hbar} \right]}}{2^N} \sum_{k=0}^{N-1} \binom{N-1}{k} e^{-\frac{2i}{\hbar} tUk} \\
& =  N \frac{e^{ i \left[\phi + (N-1) \frac{tU}{\hbar}\right]}}{2^N} \left( 1+ e^{-\frac{2i}{\hbar} tU} \right)^{N-1} \\
&= N \frac{e^{i\phi}}{2} \left[ \cos \left ( \frac{tU}{\hbar} \right)\right]^{N-1}.
\end{split}
\end{equation}

Considering the superposition of phase states~\eqref{supphase} , we have that
\begin{equation}
\label{intpha}
\begin{split}
&\bra{\phi} \hat{a}^{\dagger^k}_{L} \hat{a}_{R}^{k} \ket{\phi+\pi} = \frac{1}{N! 2^N} \sum_{l,m = 0}^{N} \binom{N}{l} \binom{N}{m} e^{i\phi(l-m)}\times \\
& (-1)^{m}  \bra{0} \hat{a}_{L}^{N-l}\hat{a}_{R}^{l}  \hat{a}^{\dagger^k}_{L} \hat{a}_{R}^{k} \left(\hat{a}^{\dagger}_{L} \right)^{N-m} \left(\hat{a}^{\dagger}_{R} \right)^m \ket{0} \\
&=\frac{1}{N! 2^N} \sum_{l,m = 0}^{N} \binom{N}{l} \binom{N}{m} e^{i\phi(l-m)} (-1)^{m} \delta_{l, k + m} l! \left(N-l+k \right)! \\
&= \frac{N!}{2^N} \sum_{l=0}^{N-k} \binom{N-k}{l} (-1)^{l-k}.
\end{split}
\end{equation}
%

With a similar computation, the time evolution of eq.~\eqref{intpha} becomes
\begin{equation}
\label{intpha2}
\begin{split}
&\bra{\phi} e^{-\frac{i}{\hbar} t U \hat{a}^{\dagger}_{L} \hat{a}_{L}\hat{a}^{\dagger}_{R}\hat{a}_{R} }\, \hat{a}^{\dagger^k}_{L} \hat{a}_{R}^{k} \, e^{\frac{i}{\hbar} t U \hat{a}^{\dagger}_{L} \hat{a}_{L}\hat{a}^{\dagger}_{R}\hat{a}_{R} }\ket{\phi+\pi} \\
&= \frac{N!}{2^k} \left(-1 \right)^k e^{i\phi k} i^{N-k} e^{-\frac{i}{\hbar} U t k^2} \left[\sin \left(\frac{U k t}{\hbar} \right) \right]^{N-k} \, .
\end{split}
\end{equation}
%

\section{CSL master equation}
\label{AppC}

The general density matrix in the two-mode approximation can be written as follows:
\begin{equation}
\label{4.numberbasis}
\hat{\rho} = \sum_{\substack{k,j=0}}^{N} c_{k,j} \left(\hat{a}^{\dagger}_{L}\right)^{N-j} \left(\hat{a}^{\dagger}_{R}\right)^j\ket{0}\bra{0} \left(\hat{a}_{L}\right)^{N-k} \left(\hat{a}_{R}\right)^k,
\end{equation}
where $ c_{k,j}$ are complex coefficients satisfying the self-adjointness and normalization conditions of the density matrix. We use the expansion eq.~\eqref{4.numberbasis} in order to solve the master equation~\eqref{csltwomode}, 
as follows:
\begin{equation}
\label{4.solvingcoefficients}
\begin{split}
& \frac{d}{dt} \bra{0} \left(\hat{a}_{L}\right)^{N-m} \left(\hat{a}_{R}\right)^m \hat{\rho}(t)\left(\hat{a}^{\dagger}_{L}\right)^{N-l} \left(\hat{a}^{\dagger}_{R}\right)^l\ket{0} \equiv \dot{\rho} (m,l,t) \\
&= \frac{iU}{\hbar} \bra{0} \left(\hat{a}_{L}\right)^{N-m} \left(\hat{a}_{R}\right)^m \left[\hat{a}^{\dagger}_{L}\hat{a}_{L}\hat{a}^{\dagger}_{R}\hat{a}_{R},\hat{\rho}(t)\right] \times \\
& \left(\hat{a}^{\dagger}_{L}\right)^{N-l} \left(\hat{a}^{\dagger}_{R}\right)^l\ket{0}  - \frac{\lambda A^2 \bar{\gamma}}{2} \times \\
&\sum_{\substack{ i= L,R}}  \bra{0} \left(\hat{a}_{L}\right)^{N-m} \left(\hat{a}_{R}\right)^m \left [\hat{a}^{\dagger}_{i}\hat{a}_i, \left[ \hat{a}^{\dagger}_{i}\hat{a}_i, \hat{\rho} (t) \right] \right]\left(\hat{a}^{\dagger}_{L}\right)^{N-l} \left(\hat{a}^{\dagger}_{R}\right)^l\ket{0} \\
&=  \frac{iU}{\hbar} \left[ \left(N-m\right)m - \left(N-l\right)l\right] \rho (m,l,t) -  \\
&\lambda A^2 \bar{\gamma} \left(m-l\right)^2  \rho (m,l,t) \\
&\Rightarrow \rho (m,l,t) = e^{ \frac{iUt}{\hbar} \left(l+m-N\right)\left(l-m\right)} e^{- \lambda A^2 \bar{\gamma}t\left(m-l\right)^2}\rho (m,l,0).
\end{split}
\end{equation}
From eq.~\eqref{4.solvingcoefficients}, the density matrix is easily obtained. 

We also have that 
\begin{equation}
\label{4.CSLphasecoherence}
\begin{split}
&\langle \hat{a}^{\dagger}_{L} \hat{a}_R \rangle^{\text{CSL}}_{t} = \text{Tr} \left[ \hat{a}^{\dagger}_{L} \hat{a}_R \hat{\rho} (t) \right] \\
&= \sum_{\substack{k=0}}^{N} \frac{1}{\left(N-k\right)!k!} \bra{0} \left(\hat{a}_{L}\right)^{N-k} \left(\hat{a}_{R}\right)^k  \hat{a}^{\dagger}_{L} \hat{a}_R \times \\
& e^{- \lambda A^2 \bar{\gamma}t\left(\overset{\longleftarrow}{\hat{a}^{\dagger}_{R}\hat{a}_{R}}-\overset{\longrightarrow}{\hat{a}^{\dagger}_{R}\hat{a}_{R}}\right)^2}\hat{\rho}_{\text{Sch}}(t)\left(\hat{a}^{\dagger}_{L}\right)^{N-k} \left(\hat{a}^{\dagger}_{R}\right)^k\ket{0} \\
&= e^{-\lambda A^2 \bar{\gamma}t} \sum_{\substack{k=0}}^{N} \frac{1}{\left(N-k\right)!k!} \bra{0} \left(\hat{a}_{L}\right)^{N-k} \left(\hat{a}_{R}\right)^k  \hat{a}^{\dagger}_{L} \hat{a}_R \times \\
&\hat{\rho}_{\text{Sch}}(t)\left(\hat{a}^{\dagger}_{L}\right)^{N-k} \left(\hat{a}^{\dagger}_{R}\right)^k\ket{0} \\
&= e^{-\lambda A^2 \bar{\gamma}t} \langle  \hat{a}^{\dagger}_{L} \hat{a}_R \rangle^{\text{Sch}}_{t}.
\end{split}
\end{equation}
%

\section{Interaction between a condensate and a thermal cloud: master equation and decoherence rates}
\label{AppD}

We derive the master equation describing the dynamics of a Bose-Einstein condensate interacting with its thermal cloud. We start by writing the Hamiltonian of the total system:
\begin{equation}
\label{hamiltonian_app}
\begin{split}
&\hat{H} = \int d\ve{x} \, \hat{a}^{\dagger} (\ve{x}) \left (-\frac{\hbar^2 \nabla^2}{2m} +V_{\text{ext}}(\ve{x}) \right) \hat{a}(\ve{x}) + \\
&\frac{g}{2} \int d\ve{x} \,  \hat{a}^{\dagger} (\ve{x}) \hat{a}^{\dagger} (\ve{x}) \hat{a}(\ve{x}) \hat{a}(\ve{x}),
\end{split}
\end{equation}
We rewrite the Hamiltonian operator in terms of the energy eigenstates $\hat{a}_{i}^{\dagger}\ket{0} = \ket{\psi}_i$ of the single particle Hamiltonian $-\frac{\hbar^2 \nabla^2}{2m} +V_{\text{ext}}(\ve{x})$:
\begin{equation}
\label{hamiltonian2}
\hat{H} = \sum_i \epsilon_i \hat{a}_{i}^{\dagger} \hat{a}_{i} + \frac{1}{2} \sum_{\substack{i,j \\ k,l}} \gamma^{i,j}_{k,l} \hat{a}_{i}^{\dagger} \hat{a}_{j}^{\dagger} \hat{a}_{k}\hat{a}_{l} \, ,
\end{equation}
where 
\begin{equation}
\label{gamma_app}
\gamma^{i,j}_{k,l} = \int d\ve{x} \, \psi_{i}^{*}\left(\ve{x}\right)  \psi_{i}^{*}\left(\ve{x}\right) \psi_{k}\left(\ve{x}\right) \psi_{l}\left(\ve{x}\right) \, . 
\end{equation} 
By splitting the condensate modes $i \in [0, \dots, J] \equiv \mathcal{C}$ from the thermal modes $i \in [J+1, \dots, +\infty[ \equiv \mathcal{N}\mathcal{C}$, the Hamiltonian operator in eq.~\eqref{hamiltonian2} becomes:
\begin{equation}
\label{hamiltonian3}
\hat{H} = \hat{H}_C + \hat{H}_{NC} + \hat{V}_{int} \, ,
\end{equation} 
where
\begin{align}
&\hat{H}_C = \sum_{i \in \mathcal{C}} \epsilon_i \hat{a}_{i}^{\dagger} \hat{a}_{i} + \frac{1}{2} \sum_{\mathcal{C}} \gamma^{i,j}_{k,l} \hat{a}_{i}^{\dagger} \hat{a}_{j}^{\dagger} \hat{a}_{k}\hat{a}_{l} \label{hbec} \, ; \\
&\hat{H}_{NC} = \sum_{i \in \mathcal{N}\mathcal{C}} \epsilon_i \hat{a}_{i}^{\dagger} \hat{a}_{i} + \frac{1}{2} \sum_{\mathcal{N}\mathcal{C}} \gamma^{i,j}_{k,l} \hat{a}_{i}^{\dagger} \hat{a}_{j}^{\dagger} \hat{a}_{k}\hat{a}_{l} \label{hnonbec} \, ; \\
& \hat{V}_{int} = \frac{1}{2} \sum_{\mathcal{C} + \mathcal{N}\mathcal{C}} \gamma^{i,j}_{k,l} \hat{a}_{i}^{\dagger} \hat{a}_{j}^{\dagger} \hat{a}_{k}\hat{a}_{l} \label{vint}\, .
\end{align}
In the interaction picture (labeled by I), an initial density matrix $\hat{\rho}_0$ evolves in time as follows:
\begin{equation}
\label{rho1}
\hat{\rho}^I (t) = \text{T}\left \{ e^{-\frac{i}{\hbar}\int d\tau \, \hat{V}_{int}^{I} (\tau)}\right \}\hat{\rho}_0\text{T}\left \{ e^{\frac{i}{\hbar}\int d\tau \, \hat{V}_{int}^{I} (\tau)}\right \} \, ,
\end{equation} 
where $\text{T}\{ \cdot \}$ refers to time ordering operations. In the weak coupling limit~\cite{breuer2002theory}, we expand the two time ordered exponentials in eq.~\eqref{rho1} up to the second order in the coupling $g$, obtaining the following expression
\begin{widetext}
\begin{equation}
\label{rho2}
\begin{split}
\hat{\rho}^I (t) = &\hat{\rho}_0 -\frac{i}{\hbar}\int d\tau \, \left [ \hat{V}_{int}^{I} (\tau), \hat{\rho}_0 \right] + \left(\frac{i}{\hbar}\right)^2 \int_{0}^{t} d\tau_1 \int_{0}^{\tau_1} d\tau_2 \, \left(\hat{V}_{int}^{I} (\tau_1) \hat{V}_{int}^{I} (\tau_2) \hat{\rho}_0  + \, \hat{\rho}_0\hat{V}_{int}^{I} (\tau_2)\hat{V}_{int}^{I} (\tau_1) \right) \\
&-  \left(\frac{i}{\hbar}\right)^2 \int_{0}^{t} d\tau_1 \int_{0}^{t} d\tau_2 \, \hat{V}_{int}^{I} (\tau_1)\hat{\rho}_0 \hat{V}_{int}^{I} (\tau_2) \, . 
\end{split}
\end{equation} 
\end{widetext}
We work in the Born-Markov approximation~\cite{breuer2002theory}, with initial state given by
\begin{equation}
\label{rho3}
\hat{\rho}_0 = \hat{\rho}_{\mathcal{C}} \frac{e^{-\beta \hat{H}_{NC}}}{ \text{Tr} \left[e^{-\beta \hat{H}_{NC}}\right]} \, . 
\end{equation} 
We are interested only in the condensate modes $\mathcal{C}$, thus in eq.~\eqref{rho2} we perform a partial trace over the non-condensate modes $\mathcal{N}\mathcal{C}$. In computing the reduced master equation for the condensate, we neglect the processes that do not conserve the energy of the system, such as $C + C \to NC + NC$ and $C + C \to NC + C$, where $C$ refers to a condensate atom, and $NC$ refers to a non-condensate atom. We thus obtain the following reduced master equation in the Schr\" odinger picture:
\begin{equation}
\label{rho4}
\frac{d\hat{\rho}_{\mathcal{C}} (t)}{dt} = -\frac{i}{\hbar} \left [\hat{H}_C,\hat{\rho}_{\mathcal{C}} (t) \right] + \sum_{i= 1,2}\widetilde{\mathcal{L}}_{t}^{(i)} \left[ \hat{\rho}_{\mathcal{C}} (t) \right] \, ,
\end{equation}
where 
\begin{align}
&\widetilde{\mathcal{L}}_{t}^{(1)} \left[\hat{\rho}_{\mathcal{C}}(t) \right] = \int d\ve{x} \int d\ve{y} \, \Lambda_{\text{loss}} \left(\ve{x},\ve{y} \right) \times  \notag \\
&\left\{ \hat{a}^{\dagger} (\ve{x}) \hat{a} (\ve{y}) \hat{\rho}_{\mathcal{C}}(t) + \hat{\rho}_{\mathcal{C}}(t) \hat{a}^{\dagger} (\ve{x}) \hat{a} (\ve{y}) - 2 \hat{a} (\ve{y}) \hat{\rho}_{\mathcal{C}}(t) \hat{a}^{\dagger} (\ve{x}) \right \} \, ; \label{loss1}\\
&\widetilde{\mathcal{L}}_{t}^{(2)} \left[\hat{\rho}_{\mathcal{C}}(t) \right] = \int d\ve{x} \int d\ve{y} \, \Lambda_{\text{dec}} \left(\ve{x},\ve{y} \right) \times \notag \\ 
&\left[\hat{a}^{\dagger} (\ve{x}) \hat{a} (\ve{x}), \left[\hat{a}^{\dagger} (\ve{y}) \hat{a} (\ve{y}),\hat{\rho}_{\mathcal{C}}(t) \right]\right] \, , \label{dec1} 
\end{align}
with damping rates given by the following quantities
\begin{align}
&\Lambda_{\text{loss}} \left(\ve{x},\ve{y} \right) = \frac{g^2}{2 \hbar^2} \text{Tr} \left[\hat{a}^{\dagger} (\ve{x})\hat{a} (\ve{x})\hat{a} (\ve{x}) \times \right. \\
& \left. \hat{a}^{\dagger} (\ve{y})\hat{a}^{\dagger} (\ve{y})\hat{a} (\ve{y})  \frac{e^{-\beta \hat{H}_{NC}}}{ \text{Tr} \left[e^{-\beta \hat{H}_{NC}}\right]} \right] \, ; \label{lamlos1} \\
&\Lambda_{\text{dec}}\left(\ve{x},\ve{y} \right) = \frac{g^2}{2 \hbar^2} \text{Tr} \left[\hat{a}^{\dagger} (\ve{x})\hat{a} (\ve{x}) \hat{a}^{\dagger} (\ve{y})\hat{a}^{\dagger} (\ve{y}) \frac{e^{-\beta \hat{H}_{NC}}}{ \text{Tr} \left[e^{-\beta \hat{H}_{NC}}\right]} \right] \, . \label{lamdec1}
\end{align}
As one can note, the damping rates introduced in eqs.~\eqref{lamlos1} and~\eqref{lamdec1} are proportional, respectively, to the three-particle correlation and the two-particle correlation of the thermal cloud. For an homogeneous thermal cloud  both quantities are strongly peeked around $\ve{x} = \ve{y}$, so we can further simplify the expressions as follows:
\begin{align}
&\Lambda_{\text{loss}} \left(\ve{x},\ve{y} \right) \approx \frac{g^2}{2 \hbar^2}\text{Tr} \left[\hat{a}^{\dagger} (\ve{0})\hat{a} (\ve{0})\hat{a} (\ve{0}) \times \right. \notag \\
& \left. \hat{a}^{\dagger} (\ve{0}) \hat{a}^{\dagger} (\ve{0})\hat{a} (\ve{0})  \frac{e^{-\beta \hat{H}_{NC}}}{ \text{Tr} \left[e^{-\beta \hat{H}_{NC}}\right]} \right] \delta \left(\ve{x} - \ve{y} \right) \, ; \label{lamlos1_app} \\
&\Lambda_{\text{dec}}\left(\ve{x},\ve{y} \right) =\frac{g^2}{2 \hbar^2} \text{Tr} \left[\hat{a}^{\dagger} (\ve{0})\hat{a} (\ve{0}) \hat{a}^{\dagger} (\ve{0}) \times \right. \notag \\ 
&\left. \hat{a}^{\dagger} (\ve{0}) \frac{e^{-\beta \hat{H}_{NC}}}{ \text{Tr} \left[e^{-\beta \hat{H}_{NC}}\right]} \right] \delta \left(\ve{x} - \ve{y} \right) \, , \label{lamdec1_app}
\end{align}
where $\delta \left(\ve{x} - \ve{y} \right)$ is the Dirac delta. The expressions in eqs.~($35$) and~($36$) are easily found in standard textbooks on cold atomic systems~\cite{pethick2002bose}. Imposing the two-mode approximation on the master equation~\eqref{rho4}, the eqs.~($30$) and~($32$) are easily obtained.

\begin{acknowledgments}
Discussions with A. Smerzi are gratefully acknowledged. The authors also thank the anonymous referee for several useful comments.
M.B. and A.B.  
acknowledge financial support from the University of Trieste (FRA 2016). 
Financial support from INFN is as well acknowledged.
\end{acknowledgments}

\bibliography{bibjj}
\end{document}